\documentclass[reprint,
superscriptaddress,
amsmath,amssymb,
aps,
pra,
]{revtex4-2}
\usepackage{graphicx}
\usepackage{dcolumn}
\usepackage{bm}
\usepackage{dsfont}
\usepackage{physics}
\usepackage{bbold}
\usepackage{color}
\usepackage{soul}
\setstcolor{red}
\usepackage{ifthen}
\newboolean{showannotations}   
\setboolean{showannotations}{false} 
\newcommand*{\added}[1]{%
  \ifthenelse{\boolean{showannotations}}{{\color{blue}#1}}{#1}%
}

\newcommand*{\removed}[1]{%
  \ifthenelse{\boolean{showannotations}}{\st{#1}}{}%
}

\newcommand*{\change}[2]{%
  \ifthenelse{\boolean{showannotations}}{{\color{red}\st{#1}}{\color{blue}#2}}{#2}%
}

\begin{document} 

\title{ Mean-field Mixed Quantum-Classical Approach for \\ Many-Body Quantum Dynamics of Exciton-Polaritons }
\author{Pritha Ghosh}%
\affiliation{Department of Chemistry, Texas A\&M University, College Station, Texas 77843, USA}
\author{Arshath Manjalingal}%
\affiliation{Department of Chemistry, Texas A\&M University, College Station, Texas 77843, USA}
\author{Sachith Wickramasinghe}%
\affiliation{Department of Chemistry, Texas A\&M University, College Station, Texas 77843, USA}
\author{Saeed Rahmanian Koshkaki}%
\affiliation{Department of Chemistry, Texas A\&M University, College Station, Texas 77843, USA}
\author{Arkajit Mandal}%
\email{mandal@tamu.edu}
\affiliation{Department of Chemistry, Texas A\&M University, College Station, Texas 77843, USA}

\begin{abstract}
In this work, we use a mixed quantum-classical (mean-field) many-body  approach for simulating the quantum dynamics of excitons and exciton-polaritons beyond the single‑excitation subspace. We combine the multitrajectory Ehrenfest approach, which propagates slow degrees of freedom classically, with the Gross-Pitaevskii method, which propagates fast degrees of freedom in a mean-field fashion. We use this {\it mean-field many-body Ehrenfest approach} to analyze how the phonon-induced dynamic disorder and the many-body interaction affect the incoherent and coherent dynamics of excitons and exciton-polaritons. We examine how the number of excitations and the strength of repulsive exciton-exciton interaction nonlinearly influence the transport, Fr\"{o}hlich scattering and decoherence. 
\end{abstract}

\maketitle
{
\section{Introduction} 
The predominant use of electronics for computing and photons for high-speed communication reflects fundamental differences in the physical properties of light and matter. Polaritons, which are part-light and part-matter, form a natural bridge between the two and may provide a versatile platform for classical and quantum computing~\cite{OpalaOME2023, KavokinNRP2022, SanvittoNM2016, XiangCR2024, KeelingARPC2020, BasovNp2025}. While their matter component allows for nonlinear interactions suitable for computation~\cite{HartmannJO2016}, their photonic component allows for efficient information transport by limiting decoherence and dephasing~\cite{XuNC2023, BalasubrahmaniyamNM2023}. These dual characteristics have been exploited in the past decade to demonstrate proof-of-principle applications in classical computing~\cite{ZasedatelevNP2019, AmoNP2010}, quantum computing~\cite{GhoshNPJQ2020, KavokinNRP2022}, quantum simulators~\cite{BerloffNM2017, RojasPRB2023, BoulierAQT2020, HartmannJO2016}, and neuromorphic computing~\cite{GhoshAQT2021, BallariniNL2020, MirekNL2021}. However, the transition from proof-of-principle demonstrations to deployable polariton technologies has been slow, at least in part, due to a lack of theoretical understanding of their dynamics at a microscopic level~\cite{KeelingARPC2020}.

Simulating exciton-polaritons is a challenging task. Exciton-polaritons formed in optical cavities involve an ensemble of cavity radiation modes interacting with a large number of electronic states and nuclear degrees of freedom (phonon)~\cite{XuNC2023, SokolovskiiNC2023, MandalCR2023, ChngNL2025}, making exact quantum dynamical propagation a formidable task. Prior theoretical works~\cite{ SokolovskiiNC2023, TichauerJCP2021, MandalCR2023, ChngNL2025, koshkakiArxiv2025, BlackhamArxiv2025, TichauerAS2023, AntoniouJPCL2020, MandalCR2023, ArnardottirPRL2020, LindoyNC2023, HoffmannJCP2020, PerezPNAS2023, SharmaJCP2024, HoffmannPRA2019, KoesslerJCP2022} utilize a combination of approximations either in the light-matter model studied or in the quantum dynamical approach employed. Multitrajectory mixed quantum-classical approaches such as multitrajectory Ehrenfest~\cite{MandalCR2023, HoffmannPRA2019, LiJCTC2019, CrespoCR2018}, the fewest-switches surface-hopping~\cite{TullyJCP1990, SubotnikARPC2016, RietzeJPCC2020, AntoniouJPCL2020, KowalewskiJPCL2016, FregoniNC2018}, and various path-integral-based approaches~\cite{MandalJCPL2019, ChowdhuryJCP2021, LiJPCL2022, DepingJCP2022} have emerged as efficient theoretical methods for propagating the intricate quantum dynamics of exciton-polariton coupled to material vibrations. However, these works, when dealing with a large number of molecular degrees of freedom, confine the dynamics within the single-excited subspace, thereby missing crucial many-body effects central to many polariton-based technologies explored to date. In contrast, works that considered these nonlinear interactions utilized the mean-field Gross-Pitaevskii (GP) approach~\cite{LingyuArxiv2024, CarusottoRMP2013, AlliluevJLTP2024, HaugPRB2014, AntonelliNL2019, NespoloPRB2019}  where nuclear degrees of freedom are not explicitly included, thereby missing microscopic dynamical insights.

We address this issue by developing a mean-field many-body Ehrenfest approach that propagates the dynamics of exciton-polaritons coupled to phonon degrees of freedom  beyond the single-excited subspace. Within this approach, the fast degrees of freedom, which include photonic and electronic degrees of freedom, are propagated in a mean-field fashion using a single-permanent ansatz, equivalent to the Gross-Pitaevskii~\cite{fazioArxiv2024, CarusottoRMP2013} (GP) approach, while the slow nuclear degrees of freedom are propagated quasi-classically~\cite{TullyFD1998} using a mean force. 

Using our approach, we investigate the decoherence and transport properties of excitons and exciton-polaritons. Our numerical simulation reveals that increasing the number of excitations nonlinearly modifies mean-square displacement (MSD), coherence lifetimes, and Fr\"{o}hlich scattering owing to an interplay of enhanced phonon-induced dynamic disorder and exciton-exciton interactions.  Although the present study employs a mean-field treatment of the many-body exciton-polariton wavefunction as well as a mixed quantum-classical treatment of the nuclear degrees of freedom, it lays a crucial foundation for developing more accurate yet computationally affordable many-body quantum-dynamical methods.




\section{Theory}
{\bf Light-Matter Hamiltonian.} In this work we consider a multimode generalized Tavis-Cummings Hamiltonian~\cite{KeelingARPC2020, MandalCR2023, MandalNL2023, SokolovskiiNC2023} with the matter (excitonic) part described with a Bose-Hubbard-Holstein Hamiltonian~\cite{KeelingARPC2020,CarusottoRMP2013}. Here, following previous works~\cite{TichauerJCP2021, XuNC2023, ChngNL2025, ArnardottirPRL2020}, we consider a 2D world where two perfectly reflective mirrors run parallel to the $\hat{x}$ axis and are located at  $\pm L_y/2$,  such that the cavity quantization direction is $\hat{y}$. Further, we place a one-dimensional excitonic material consisting of $N$ sites midway between the two mirrors; its $n$-th lattice site sits at $r_n = n\,a\,\hat{x}$ (for $n\in[0,1,\dots, N-1]$), where $a$ is the lattice constant, so the chain lies entirely along the $\hat{x}$ direction. Further, we employ a periodic boundary condition along the $\hat{x}$ direction such that $r_{N} = r_0$ and $(N-1) a = L_x$. 

The light-matter Hamiltonian in atomic units is written as
\begin{equation}\label{Hamiltonian-LM}
\hat{H}_\mathrm{LM} = \hat{H}_\mathrm{e} + \hat{H}_\mathrm{p} + \hat{H}_\mathrm{e-p} + \hat{H}_\mathrm{c} + \hat{H}_\mathrm{e-c}. 
\end{equation}
Here $\hat{H}_\mathrm{e}$ is the excitonic Hamiltonian that is modeled as an interacting Bose gas~\cite{KeelingARPC2020, CombescotEPJB2009} and is described using a Bose-Hubbard Hamiltonian~\cite{KeelingARPC2020, ByrnesPRB2010} written as
\begin{equation}
\hat{H}_\mathrm{e} = \sum_{n, m} \epsilon_{n,m} \hat{X}_{n}^\dagger \hat{X}_{m} +  \frac{U }{2}\sum_{n} \hat{X}_n^\dagger \hat{X}_n^\dagger \hat{X}_n \hat{X}_n,
\end{equation}
where $\hat{X}_{n}^{\dagger}$ is a bosonic creation operator~\cite{Agranovich1968SPJ} that creates an exciton at a site $n$. Here, $\epsilon_{nm} = \epsilon_{0}\delta_{nm} - \tau (\delta_{n, n+1} + \delta_{n, n-1})$ are the one-body parameters, with $\epsilon_{0}$ as the on-site energy and $\tau$ as a hopping parameter. Finally, $U$ determines the on-site exciton-exciton interaction. 

The phonon Hamiltonian $\hat {H}_\mathrm{p}$ simply describes a set of harmonic oscillators and is written as
\begin{equation}
\hat{H}_\mathrm{p} = \sum_{n} \frac{\hat P_n^2}{2} + \frac{1}{2}\omega^2{\hat R_n^2}, 
\end{equation}
where $\{\hat P_n\}$ and $\{\hat R_n\}$ are the momentum and position operators corresponding to the phonon degrees of freedom that couple to the excitons. In this work, we adopt the Holstein form~\cite{XuNC2023, JankeJCP2020, Wang2011JCP, ChengJCP2008, ChngNL2025} of the exciton–phonon coupling, written as
\begin{equation}\label{ep-coupling}
\hat{H}_\mathrm{e-p}= \gamma \sum_{n}  \hat{X}_{n}^\dagger\hat{X}_n  \hat{R}_n ,
\end{equation}
where $\gamma$ is the exciton–phonon coupling. The multimode cavity Hamiltonian $\hat{H}_\mathrm{c}$ is written as~\cite{MandalCR2023, KeelingARPC2020, ArnardottirPRL2020}
\begin{equation}
\hat{H}_\mathrm{c} =  \sum_{k} \hat{a}^\dagger_{k}\hat{a}_{k} \omega_k, 
\end{equation}

where $\hat{a}_k^{\dagger}$ creates a photon of frequency $\omega_k = \frac{c}{\eta}\sqrt{k^2 +{\pi/L_y}}$, with ${\eta} = 2.4$, the refractive index of the material and $c \approx 137$ a.u., the speed of light. Here, $k$ is the in-plane wave vector, and ${\pi/L_y}$ is the transverse component, corresponding to the fundamental cavity mode. 

Finally, $\hat{H}_\mathrm{e-c}$  describes the exciton-photon interactions and is written as
\begin{equation}
\hat{H}_\mathrm{e-c} =  \sum_{n,k} \frac{ \Omega_k}{\sqrt{N}}\bigg[\hat{a}^{\dagger}_{k}\hat{X}_{n}e^{-i k  \cdot r_{n}} + \hat{a}_{k}\hat{X}_{n}^{\dagger}e^{i k \cdot r_{n}} \bigg].
\end{equation}
where $\Omega_k = \Omega_0 \sqrt{\frac{\omega_0}{\omega_k}}$ with $ \Omega_0 $ as light-matter coupling strength and $N = 30001$ is the number of sites chosen here (unless otherwise mentioned). We consider the number of cavity modes to also be $N$, although only $\approx$ 1000 cavity modes are sufficient to obtain converged results.

{\bf Many-body Ehrenfest Approach.} Within the multitrajectory Ehrenfest approach, the nuclear degrees of freedom are treated classically ($\{\hat{R}_n, \hat{P}_n\} \rightarrow \{{R}_n, {P}_n\} $). Meanwhile, the excitonic-photonic part is propagated quantum mechanically. Despite this mixed quantum-classical treatment, it is still prohibitively expensive to propagate the many-body exciton-photon wavefunction $|\Psi\rangle$ using the time-dependent Schr\"{o}dinger equation. To address this challenge, we use the following time-dependent single permanent ansatz, formally equivalent to the mean-field Gross-Pitaevskii approach~\cite{fazioArxiv2024, CarusottoRMP2013} for the exciton-photon wavefunction confined in the $N_\mathrm{ex}$th excitation subspace, written as 
\begin{equation}\label{ansatz}
|\Psi(t)\rangle = \frac{\big(\hat {B}_{0}^{\dagger}(t)\big)^{N_\mathrm{ex}}}{\sqrt{N_\mathrm{ex}!}}|\bar{0}\rangle,
\end{equation}
where \( |\bar{0}\rangle \) represents the vacuum state and $\hat {B}_0^{\dagger}(t)$ is a bosonic creation operator, defined as a linear combination of excitonic and photonic operators, and written as 
\begin{align}\label{B0}
 {B}_0^{\dagger}(t) = \sum_{n} \psi_n(t) \cdot \hat{X}^{\dagger}_n +  \sum_{k} \phi_k(t) \cdot \hat{a}^{\dagger}_k,
\end{align}
where $\psi_n(t)$ ($\phi_k(t)$) are the exciton (photon) amplitude in real (reciprocal) space at time $t$. Throughout this work, we adopt the dilute Bose‑gas regime, with $N_\mathrm{ex}/N \ll 1$.

Following the standard Dirac–Frenkel time-dependent variational principle, which leads to  $\frac{\partial}{\partial \xi^*}\left(\langle\Psi(t)|\hat{H}-i\frac{\partial}{\partial t}|\Psi(t)\rangle\right)=0$ where $\xi^* \in \{ \psi_n^*(t), \phi_k^*(t) \}$, we obtain the equation of motion of $\psi_n(t)$ and $\phi_k(t)$ as
\begin{align}\label{electronic-DOF}    
 i\dot{\psi_n}(t) &=  \epsilon_{0} \psi_{n}(t) -\tau (\psi_{n+1}(t) + \psi_{n-1}(t))  \nonumber \\
 &+ \gamma \psi_{n}(t) {R}_{n}(t) +\sum_{k} \frac{\Omega_k}{\sqrt{N}} \phi_{k}(t)e^{-i k \cdot r_{n}} \nonumber \\
 &+{U}(N_\mathrm{ex}-1)|\psi_n(t)|^2\psi_n (t),  \\
  i\dot{\phi_k}(t) &=  \phi_{k}(t) \omega_{k} +  \sum_{n} \frac{\Omega_k}{\sqrt{N}} \psi_{n}(t)e^{-i k \cdot r_{n}}.
\end{align}
In this work, we use a split-operator technique to efficiently propagate $\psi_n(t)$ and $\phi_k(t)$
with details provided in the Supporting Information (SI). Finally,  we find the nuclear equation of motion using the mean-field ansatz in Eq.~\ref{ansatz}, written as 
\begin{align} \label{nuclear-EOM}
\dot{R}_n(t) &= P_n(t) \\
\dot{P}_n(t) &= -\omega^2R_n - \Big\langle \Psi(t)\Big|\frac{d \hat{H}_\mathrm{e-p}}{dR_n}\Big|\Psi(t)\Big\rangle \nonumber \\
&= -\omega^2R_n - \gamma |\psi_{n}(t)|^2 N_\mathrm{ex}.
\end{align}

As illustrated by Eqs.~\ref{nuclear-EOM} and \ref{electronic-DOF}, the exciton–phonon interaction in Eq.~\ref{ep-coupling} does not mix different excitation subspaces and therefore preserves the total number of excitations \(N_\mathrm{ex}\).  

The initial state of our system is $\hat\rho(0)  = |\Psi (0)\rangle \langle \Psi (0)|\otimes \hat \rho_R$ where $|\Psi (0)\rangle$ is the initial many-body wavefunction that describes the excitonic-photonic subsystem and $\hat\rho_R$ is the initial nuclear density, that is, the canonical thermal density $\hat\rho_R = e^{-\beta \sum_n\hat{P}^2_n/2 + \omega^2\hat R^2_n/2}$. The initial nuclear positions and momenta $\{R_{n}(0), P_n(0)\}$ are sampled from the Wigner distribution $[\hat\rho_R]_W \propto e^{-\tanh(\beta\omega) \sum_n [\omega R^2_n  + P^2_n/\omega] }$. The time-dependent reduced density matrix of the exciton-photon subsystem is then computed by averaging over nuclear trajectories, $\hat\rho_s(t) =\frac{1}{N_\mathrm{T}} \sum_j |\Psi_j (t)\rangle \langle \Psi_j (t)|$, where $N_\mathrm{T}$ is the total number of trajectories and $j$ runs over different trajectories.

{\bf Computing Purity.} In this work we are interested in the cavity-modified decoherence beyond the single-excited subspace. We find a numerically convenient approach to computing purity $\mathcal{P}(t)$, that quantifies coherence of a density matrix and is defined as

\begin{align}\label{purity1}
\mathcal{P}(t) = \mathrm{Tr}[\hat\rho_s^2(t)] = \frac{1}{N_\mathrm{T}^2}\sum_{i,j} \big|\langle \Psi_i (t)|\Psi_j (t)\rangle\big|^2. 
\end{align}

Here, $|\Psi_j (t)\rangle = \frac{1}{\sqrt{N_\mathrm{ex}!}}[\hat B^\dagger_{0j}(t)]^{N_\mathrm{ex}} |\bar{0}\rangle$, where we introduced the index $j$ to $\hat B^\dagger_{0j}$ to distinguish between different trajectories. Next, we express $\hat B^\dagger_{0j}(t) = A_{ij}^{[0]} \hat B^\dagger_{0i}(t) + A_{ij}^{[1]} \hat B^\dagger_{1i}(t) + ...$, where $B^\dagger_{\alpha i}(t)$  with $\alpha \ge 1$  commutes with $\hat B_{0i}(t)$, that is, $[\hat B^\dagger_{0i}(t), \hat B_{\alpha i}(t)] = 0$. We compute the overlap in Eq.~\ref{purity1} as

\begin{align} 
\langle \Psi_i (t)|\Psi_j (t)\rangle 
&= \langle \bar{0}| \frac{[\hat B_{0i}(t)]^{N_\mathrm{ex}}}{\sqrt{N_\mathrm{ex}!}} \cdot \frac{[ A_{ij}^{[0]}\hat B^\dagger_{0i}(t) + ...]^{N_\mathrm{ex}}}{\sqrt{N_\mathrm{ex}!}} |\bar{0}\rangle \nonumber \\
&= (A_{ij}^{[0]})^{N_\mathrm{ex}},
\end{align}
where $A_{ij}^{[0]} = \sum_{n}{\psi}_{n, i}^*(t) \cdot  {\psi}_{n, j}(t) + \sum_{k}{\phi}^*_{k,i}(t) \cdot  {\phi}_{k,j}(t)$  with  $j$ as the trajectory index. Thus, we compute purity as

\begin{align} 
\mathcal{P}(t)  = \frac{1}{N_\mathrm{ex}} + \frac{2}{N_\mathrm{ex}^2}\sum_{i>j} \big|A_{ij}^{[0]}\big|^{2N_\mathrm{ex}}.
\end{align}

This expression enables us to efficiently compute the purity of the many-body exciton-photon subsystem without the need to directly compute $\hat \rho_s(t)$, which is prohibitively expensive.

\begin{figure*}[t]
    \centering
    \includegraphics[width=1\linewidth]{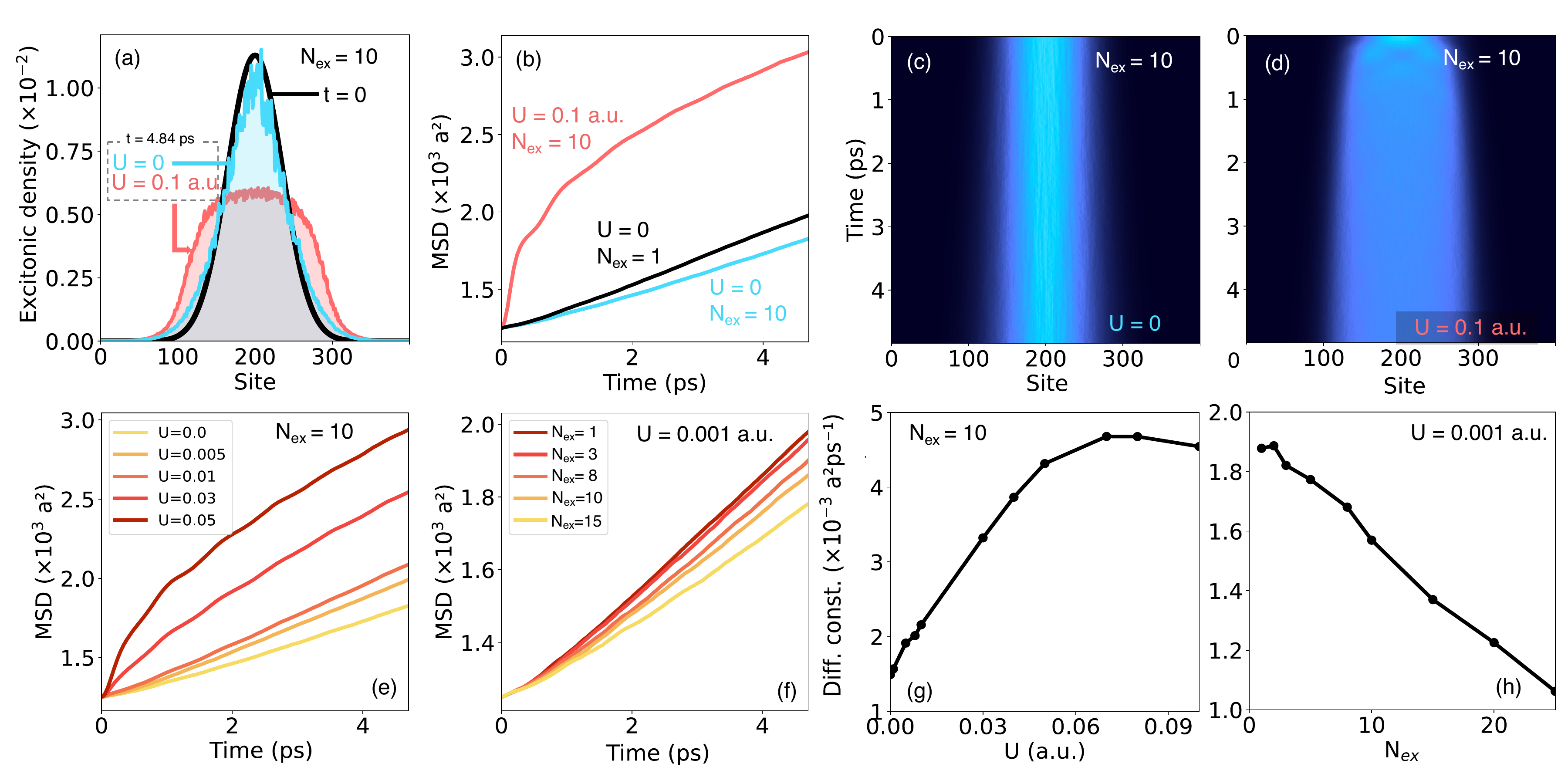}
    \caption{Quantum dynamics of a bare excitonic material. (a)  Initial excitonic density (at $t=0$) (black solid line), excitonic density at $t=4.84$ ps for non-interacting excitons(blue solid line), and for interacting excitons(red solid line). (b) Time-dependent mean-square displacement (MSD) ($\mathrm{\AA}^{2}\mathrm{ps^{-1}}$) for single exciton (black solid line), 10 non-interacting excitons (blue solid line), and 10 interacting ($U=0.1$ a.u.) excitons (red solid line). (c) Excitonic density evolution for 10 non-interacting excitons. (d) Excitonic density evolution for 10 interacting excitons. (e) Time-dependent MSD for different variations of on-site interaction strength $U$ with a fixed number of excitations ($N_\text{ex}=10$). (f) Time-dependent MSD, for different numbers of excitations at constant on-site interaction strength $U=0.001$ a.u. (g) Diffusion constant ($\mathrm{\AA}^{2}\mathrm{ps^{-1}}$), over different variations of interaction strength $U$ at constant number of excitations, (${N_\text{ex}=10}$). (h) Diffusion constant over different variations of number of excitations (${N_\text{ex}}$) at fixed interaction strength ($U=0.001$ a.u.). All simulations have been performed with hopping parameter $\tau = 300 $ cm$^{-1}$ and phonon-coupling strength $\gamma=3500$ cm$^{-1}\mathrm{\AA}^{-1}$, at temperatures $150$ K.} 
    \label{Fig1}
\end{figure*}

\section{Results and Discussion}
{\bf Exciton Dynamics.} Fig.~\ref{Fig1} presents the dynamics of a bare excitonic material, for a linear (1D) lattice of 400 sites.
Here, we have used 500 trajectories to obtain converged excitonic dynamics. The bare excitonic Hamiltonian is written as,
\begin{equation}
\hat{H}_\mathrm{ex} = \hat{H}_\mathrm{e} + \hat{H}_\mathrm{p} + \hat{H}_\mathrm{e-p} .
\end{equation}

  \begin{figure*}[t]
    \centering
    \includegraphics[width=1.0\linewidth]{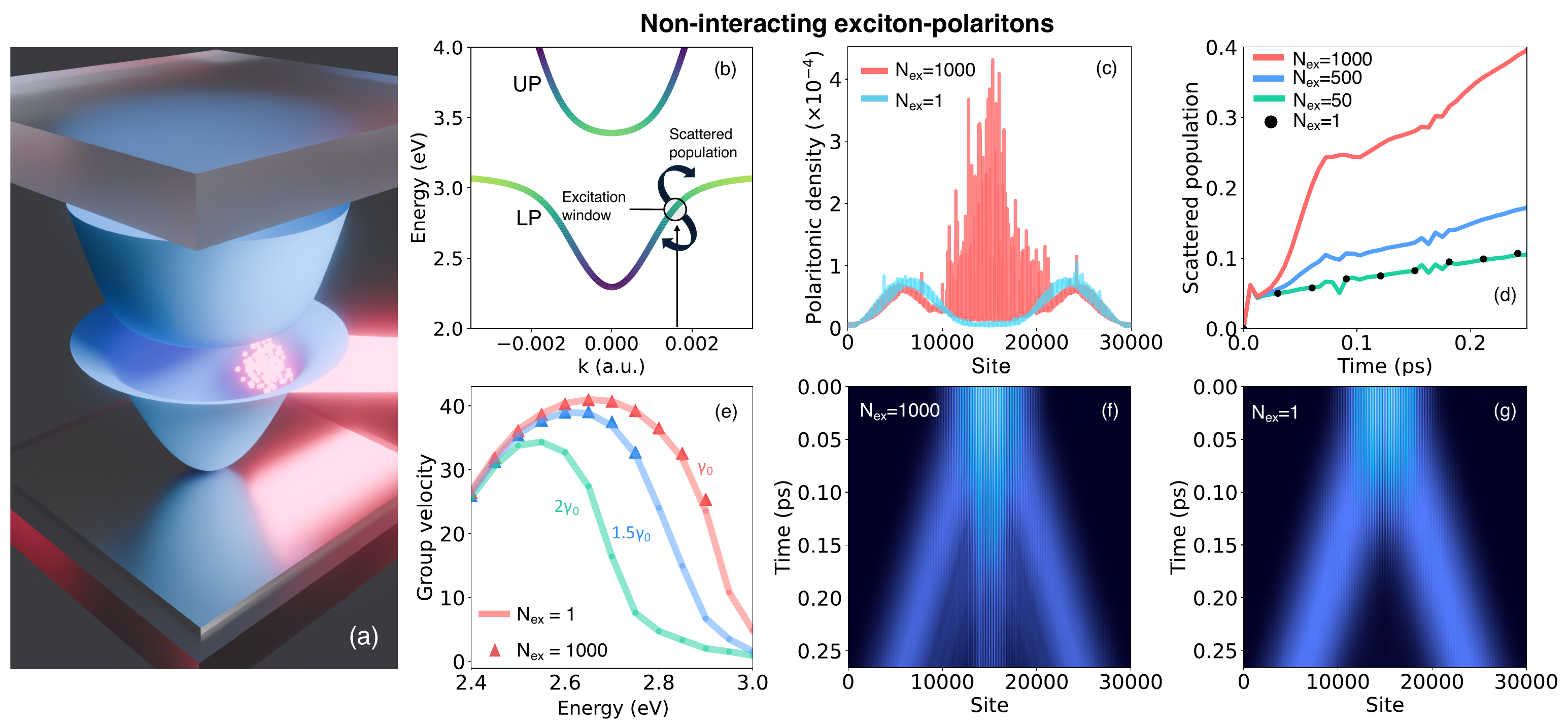}
    \caption{Dynamics of  non-interacting exciton-polaritons in the strong coupling regime. (a) Schematic illustration of polariton production by optimal excitation of material through laser in the lower polariton band. (b) Upper and lower polariton bands in the exciton-polariton dispersion plot, energies recorded in electron volts (eV), curved arrows denote Fr\"ohlich scattering. (c) Polaritonic population  at $t \approx 0.25$ ps under single excitation (blue lines) and multiple-excitations ($N_\text{ex}=1000$ ) (red lines). (d) Fr\"ohlich scattering in reciprocal space as a function of time for gradually increasing number of excitations. (e) Group velocity  obtained for both single (solid lines) and multiple excitations (triangles), with phonon-coupling parameter $\gamma=\gamma_{0} $ (red) $\gamma= 1.5\gamma_{0} $ (blue) and $\gamma= 2\gamma_{0} $ (green). (f) Polariton density evolution for $N_text{ex}=1000$. (g) Polariton density evolution for single excitation. The temperature is set to 300K. } 
    \label{Fig2}
\end{figure*}

In Fig.~\ref{Fig1}a-d, we compare the excitonic dynamics in the presence and absence of the on-site interaction term, \(U\), and with $N_\mathrm{ex} = 10$. Fig.~\ref{Fig1}a presents the site-resolved (relative) excitonic density, defined here as $\frac{1}{N_\mathrm{ex}}\big\langle \langle\Psi (t)|\hat{X}_n^{\dagger} \hat{X}_n| \Psi (t)\big\rangle_\mathrm{MFE}= \big\langle |\psi_{n}(t)|^2 \big\rangle_\mathrm{MFE}$, where $\big\langle ... \big\rangle_\mathrm{MFE} $ represents trajectory averaging, which we perform at the time of \(t\approx4.8 \) ps for an initially localized gaussian density (black solid line in  Fig.~\ref{Fig1}a). The initial density,  which is centered in the middle ($n = N/2$) of the lattice, is described via the coefficients $\psi_{n}(t = 0) = \frac{1}{\sqrt{\mathcal{N}}} \exp[{-\frac{(n-N/2)^2}{2\sigma^2}}]$, with  $\sigma = 50 $ being the width of the distribution, and $\sqrt{\mathcal{N}}$ is a normalization constant.

In the absence of on-site interactions ($U = 0$),  $N_\mathrm{ex}$ only modifies nuclear back-reaction in Eq.~\ref{nuclear-EOM}, effectively enhancing dynamic disorder. We note, however, that an increase in $N_\mathrm{ex}$ cannot simply be viewed as an increase in the exciton-phonon coupling $\gamma$. This is because while $N_\mathrm{ex}$ linearly  scales $\gamma$ in the nuclear equation of motion in Eq.~\ref{nuclear-EOM}, $\gamma$ in the electronic equation of motion in Eq.~\ref{electronic-DOF} remains unaltered. We observe that an increase in $N_\mathrm{ex}$, in the absence of an on-site interaction, leads to slower diffusion due to higher dynamical disorder. To check this, we compute the time-dependent MSD directly from the excitonic density as
\begin{align}\label{MSD}
    \mathrm{MSD} = \sum_n \left(r_n - \frac{N\,a}{2}\right)^2 \big\langle |\psi_{n}(t)|^2 \big\rangle_\mathrm{MFE}.
\end{align}
Fig.~\ref{Fig1}b illustrates that the MSD at $N_\mathrm{ex} = 10$ and $U = 0$ (blue solid line) reaches a lower value in comparison to the single-excitation scenario with $N_\mathrm{ex} = 1$ (black solid lines). Overall, at $N_\mathrm{ex} = 10$ and  $U = 0$, the excitonic density exhibits slower diffusive motion (time dependence of MSD is linear) with the density remaining a gaussian at longer times, as shown in  Fig.~\ref{Fig1}a (solid blue line). This is further illustrated in ~\ref{Fig1}c, which presents the time-dependent excitonic density at $N_\mathrm{ex} = 10$ and $U = 0$.

At the same time, in the presence of on-site interactions, \(U=0.1 \) a.u., the excitonic density exhibits sub-diffusive dynamics, with much faster ballistic propagation  at shorter times. The MSD in this scenario, presented in Fig.~\ref{Fig1}b (red solid line), increases quadratically at shorter times ($t < 100$ fs). At longer times, on-site interactions produce a pronounced flattening of the excitonic density (see the red solid line in Fig.~\ref{Fig1}a). This is because the accumulation of population over a single site is energetically unfavorable due to the presence of the term ${U}(N_\mathrm{ex}-1)|\psi_n(t)|^2$ in Eq.~\ref{electronic-DOF}. This nonlinear excitonic dynamics in the presence of on-site interactions is further illustrated in Fig.~\ref{Fig1}d, which presents the time-dependent excitonic density at \(U=0.1 \) a.u. and $N_\mathrm{ex} = 10$.


Fig.~\ref{Fig1}e presents the MSD at various on-site interaction strengths. We find that at lower on-site interaction with $U \lessapprox 0.01$ a.u., the excitonic density propagates faster (in comparison to $U = 0$) but remains diffusive, which originates from the repulsive nature of the on-site interaction,  inducing an extra hopping in addition to $\tau$, thus, allowing a faster diffusion. The diffusion constant, defined here as $D = \lim_{t\rightarrow t_\infty}\frac{1}{2}\frac{d}{dt} \mathrm{MSD}$ plotted against the on-site interaction $U$, is presented in Fig.~\ref{Fig1}g. Here we set $t_\infty = 4$ ps. The diffusion constant steadily rises at low on-site interaction strengths and saturates at around $U\approx 0.08$ a.u. We do not report diffusion constants at higher values of the on-site interaction, as the overall excitonic dynamics become sub-diffusive. 

Fig.~\ref{Fig1}f presents the MSD at various $N_\mathrm{ex}$ with the corresponding diffusion constants presented in Fig.~\ref{Fig1}h for small on-site interaction. Overall, we find that $N_\mathrm{ex}$ monotonically decreases the diffusion constant, which corroborates  our understanding of excitation-enhancement of dynamical disorder. Similar results are found when completely switching off $U$ (not shown here). 


{\bf Exciton-Polariton Dynamics.} Fig.~\ref{Fig2} shows the dynamics of non-interacting (\(U=0\)) exciton–polaritons in an optical cavity, illustrated in Fig.~\ref{Fig2}a. The total Hamiltonian of the exciton-polariton system is provided in Eq.~\ref{Hamiltonian-LM}. To accurately  capture the rapid exciton–polariton transport, we choose \(N=30001\), which corresponds to a lattice length of approximately \(36\,\mu\text{m}\) for a lattice constant \(a = 12\,\text{\AA}\). Here, we have used $100$ trajectories to obtain converged exciton-polariton dynamics (Fig.~\ref{Fig2}-\ref{Fig4}), while for purity presented in Fig.~\ref{Fig5}  we have used $30$ trajectories.

Fig.~\ref{Fig2}b presents the polariton dispersion computed by diagonalizing the pure exciton-polariton Hamiltonian in the absence of on-site interactions and phonon couplings. The pure exciton-polariton Hamiltonian is written as

\begin{align}\label{EP-Hamiltonian}
    \hat{H}_\mathrm{LM} &- \hat{H}_\mathrm{p} - \hat{H}_\mathrm{e-p} - \frac{U }{2}\sum_{n} \hat{X}_n^\dagger \hat{X}_n^\dagger \hat{X}_n \hat{X}_n \nonumber \\
    &= \sum_k \text{\raisebox{1ex}{$\begin{bmatrix}
        \hat{a}_k^\dagger & \hat{X}_k^\dagger 
    \end{bmatrix}$}}
    \begin{bmatrix}
    \omega_k  & \Omega_k \\
     \Omega_k & \epsilon_k
      \end{bmatrix} \begin{bmatrix}
        \hat{a}_k \\ \hat{X}_k 
    \end{bmatrix} \nonumber \\
    &= \sum_k \left(\hat{P}_{k,+}^\dagger \hat{P}_{k,+} \omega_{+,k}  + \hat{P}_{k,-}^\dagger \hat{P}_{k,-} \omega_{-,k}\right)
\end{align} 
where $ \epsilon_k =  \epsilon_0 - 2\tau \cos(k\cdot a)$ is the excitonic dispersion relation and $\hat{P}_{\pm,k}^{\dagger}$ create an upper/lower polariton at wave vector $k$ in the reciprocal space. Similarly, $\hat{X}_{k}^{\dagger} = \frac{1}{\sqrt{N}} \sum_k e^{i k \cdot r_n} \hat{X}_n$ creates an exciton with a wave vector $k$. The polaritonic bands $\omega_{\pm,k}$ presented in Fig.~\ref{Fig2}b, are obtained by diagonalizing the $2\times 2$ matrix in Eq.~\ref{EP-Hamiltonian}. Here we set the Rabi splitting at $k = 0$, $\Omega_0 = 0.483 $ eV. In the following, we analyze the exciton-polariton dynamics in the lower polariton. 

To study the exciton-polariton dynamics, we consider an initial state localized in an energy window $E_0 \pm \Delta E_0$ where $\Delta E_0\approx 0.0125$ eV chosen here. This energy window defines a lower polaritonic subspace spanned by $\mathcal{K} \equiv \{\hat{P}_{\kappa,-}\}$ where $E_0 -  \Delta E_0 < \omega_{-,\kappa} < E_0 + \Delta E_0$. The initial many-body exciton-polariton wavefunction is then written as (also see Eq.~\ref{ansatz})
\begin{align}
|\Psi(0)\rangle &= \frac{\Big( \sum_{k\in \mathcal{K}} \xi_k(0) \hat{P}_{-,k}^\dagger  \Big)^{N_\mathrm{ex}}}{\sqrt{N_\mathrm{ex}!}}|\bar{0}\rangle \nonumber \\
&= \frac{\Big( \sum_{k\in \mathcal{K}} [\psi_k(0) \hat{X}_k^\dagger  + \phi_k(0) \hat{a}_k^\dagger] \Big)^{N_\mathrm{ex}}}{\sqrt{N_\mathrm{ex}!}}|\bar{0}\rangle,
\end{align}
where $\psi_k(0)$ is the Fourier transform of $\psi_n(0)$, and $\{ \xi_k(0)\}$ are obtained by minimizing (via a Monte Carlo approach, see SI) the MSD defined in Eq.~\ref{MSD}, to obtain a spatially localized exciton-polariton wavefunction.

Fig.~\ref{Fig2}c presents the polariton density at time $t \sim 250$ fs in the absence of on-site interactions ($U = 0$). Here we define  the polaritonic density  as $\frac{1}{N_\mathrm{ex}}\big\langle \langle\Psi (t)|(\hat{X}_{n}^{\dagger} \hat{X}_{n} + \hat{a}_{n}^{\dagger} \hat{a}_{n})| \Psi (t)\big\rangle_\mathrm{MFE}= \big\langle |\phi_{n}(t)|^2  \big\rangle_\mathrm{MFE} + \big\langle |\psi_{n}(t)|^2  \big\rangle_\mathrm{MFE}$ where $\hat{a}_{n}^{\dagger} = \frac{1}{\sqrt{N}}\sum_{k}e^{-ik\cdot r_n}\hat{a}_{k}^{\dagger}$. At $N_\mathrm{ex} = 1$, the polariton density is made up of two wavelets (solid blue line) moving in opposite directions. This ballistic motion occurring even in the presence of phonons at room temperature has been extensively investigated both theoretically~\cite{TichauerAS2023, TichauerJCP2021, BlackhamArxiv2025, koshkakiArxiv2025, ChngNL2025} and experimentally~\cite{BalasubrahmaniyamNM2023, XuNC2023, PandyaAdvS2022}. At relatively high $N_\mathrm{ex} = 1000$  (note excitation per site is still very low, $N_\mathrm{ex}/N = 0.033 \ll 1$) we see additional density appearing in the center. The same can be observed in Fig.~\ref{Fig2}f-g, which presents the time-dependent polaritonic density at $N_\mathrm{ex} = 1$ and $1000$. We find that this is due to increased Fr\"{o}hlich scattering induced via the enhanced phonon-induced dynamic disorder, where the density escapes the initial excitation window $\mathcal{K}$.

\begin{figure*}[t]
    \centering
     \includegraphics[width=1.0\linewidth]{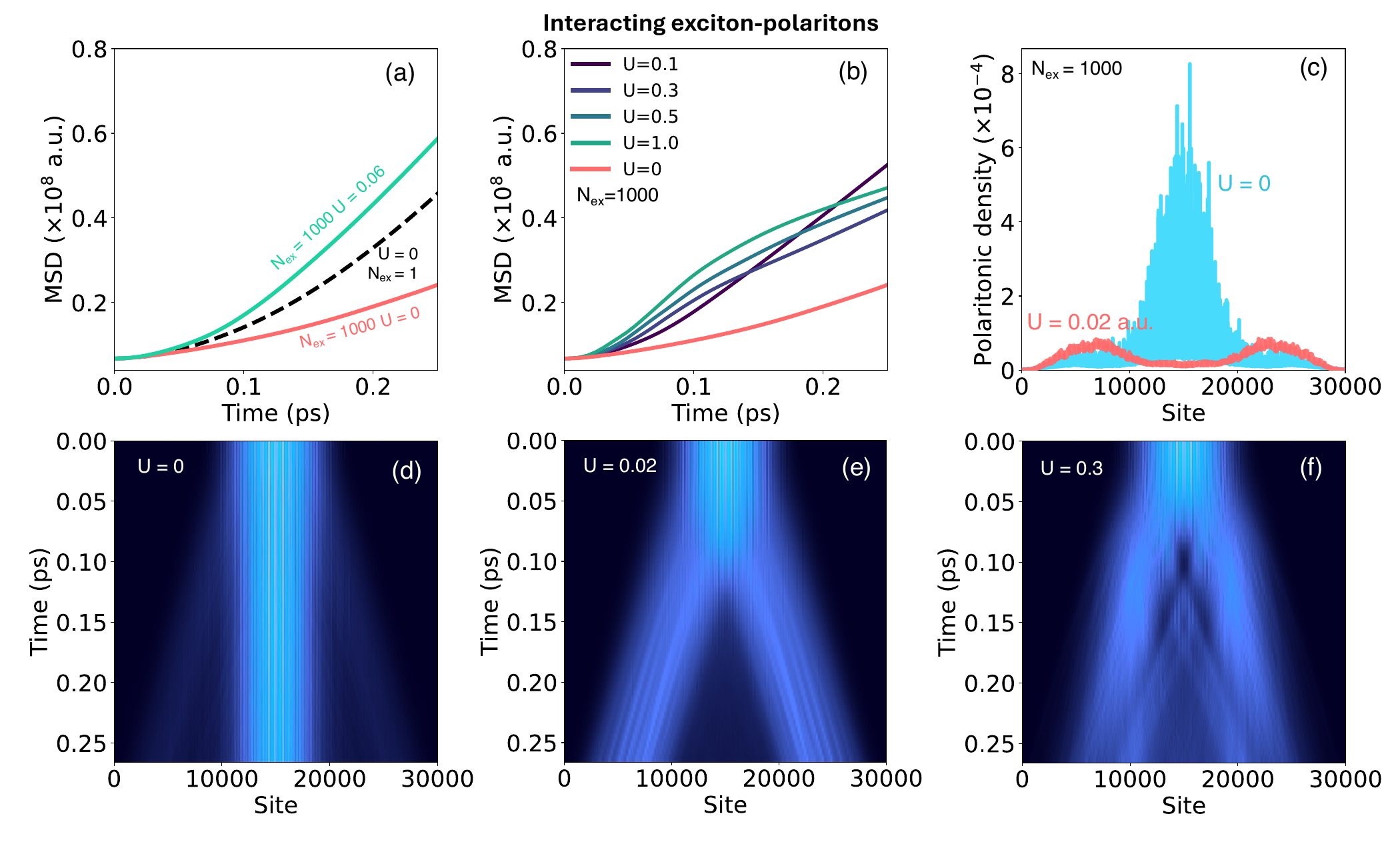}
    \caption{Dynamics of interacting excitons-polaritons. (a) Time-dependent mean-square displacement (MSD)  for $N_\text{ex}=1$ (black dashed line), $N_\text{ex}=1000$, non-interacting excitons (red solid line), $N_\text{ex}=1000$ interacting excitons (green solid line), with strength $U=0.06$ .a.u. (b) Time-dependent MSD  for fixed number of excitations ($N_\text{ex}=1000$) and different on-site interactions. (c) Polariton density at $t\approx 0.25$ ps with on-site interaction (red solid line) and without on-site interaction (blue solid line). (d) Polariton density evolution with no on-site interactions. (e) Polariton density evolution with moderate on-site interaction strength $U=0.02$  a.u. (f) Polariton density evolution with high on-site interaction strength $U=0.3$ a.u. The temperature is set to 300K.} 
    \label{Fig3}
\end{figure*} 

Fig.~\ref{Fig2}d examines phonon-induced Fr\"{o}hlich scattering as a function of the total excitation number, $N_\mathrm{ex}$.  
The scattered polariton population is obtained by summing the density that leaks outside the excitation window,
\begin{align}
\sum_{k\notin\mathcal K}
\Bigl[
\bigl\langle|\phi_k(t)|^{2}\bigr\rangle_{\mathrm{MFE}}
+
\bigl\langle|\psi_k(t)|^{2}\bigr\rangle_{\mathrm{MFE}}
\Bigr],
\end{align}
where the reciprocal-space amplitude $\psi_k(t)$ is the Fourier transforms of the real-space counterpart $\psi_n(t)$.  
At low excitation density ($N_\mathrm{ex}\!\approx\!50$), the time evolution of the scattered population and the overall dynamics remain virtually identical to the single-excitation limit; the green solid curve ($N_\mathrm{ex}=50$) and the black dots ($N_\mathrm{ex}=1$) in Fig.~\ref{Fig2}d almost completely overlap. At the same time, at much higher excitation, $N_\mathrm{ex} = 500$ and 1000, the scattering becomes progressively more pronounced, which indicates an increase in Fr\"ohlich scattering at higher number of excitation.


Fig.~\ref{Fig2}e displays the group velocity extracted from the leading wavefront of the polariton density.  
We locate the instantaneous wavefront position $n_w\cdot a$ by requiring
\[
\int_{0}^{n_w}\! dn\;
\Bigl[
\langle|\phi_n(t)|^{2}\rangle_{\mathrm{MFE}}
+
\langle|\psi_n(t)|^{2}\rangle_{\mathrm{MFE}}
\Bigr]
=w,
\]
where $a$ is the lattice constant and the threshold is fixed at $w=0.02$, corresponding to 2\% of the total polaritonic density.  
The group velocity is then defined as 
\[
v_g= \lim_{t\rightarrow t_f}\dot n_w(t)\,a,  
\]
where $t_f\approx 250$ fs in this work. Recent theoretical and experimental studies~\cite{XuNC2023, PandyaAdvS2022, BalasubrahmaniyamNM2023, BlackhamArxiv2025, YingArxiv2024} have shown that increased phonon coupling renormalizes (reduces) the polariton group velocity. Our numerical results demonstrate that higher excitation, which increases dynamic disorder effect via the scaling of \(\gamma\) in Eq.~\ref{nuclear-EOM}, does not affect the polaritonic group velocity, owing to the absence of \(N_{\mathrm{ex}}\) in the electronic-photonic equation of motion in Eq.~\ref{electronic-DOF}. Therefore, for non-interacting polaritons, \(N_{\mathrm{ex}}\) serves only to enhance phonon-induced decoherence via Fr\"{o}hlich scattering. This behavior is illustrated in Fig.~\ref{Fig2}e. At \(\gamma = \gamma_0\) (with \(\gamma_0 = 1.46\times10^{-4}\) a.u.), polaritons exhibit ballistic transport for \(E_0 < 2.9\) eV over the \(\sim250\) fs time window considered. Similarly, at higher phonon couplings, \(\gamma = 3\gamma_0/2\) the energy range where ballistic motion is observed shifts to the left to  \(E_0 < 2.8\) eV.  Finally, at \(\gamma = 2\gamma_0\) polaritons become fully diffusive (incoherent) at $N_\mathrm{ex} = 1000$ due to the enhanced decoherence discussed above and thus no group velocities can be assigned at this particular combination of $N_\mathrm{ex}$ and $\gamma$.

\begin{figure}[t]
    \centering
    \includegraphics[width=1.0\linewidth]{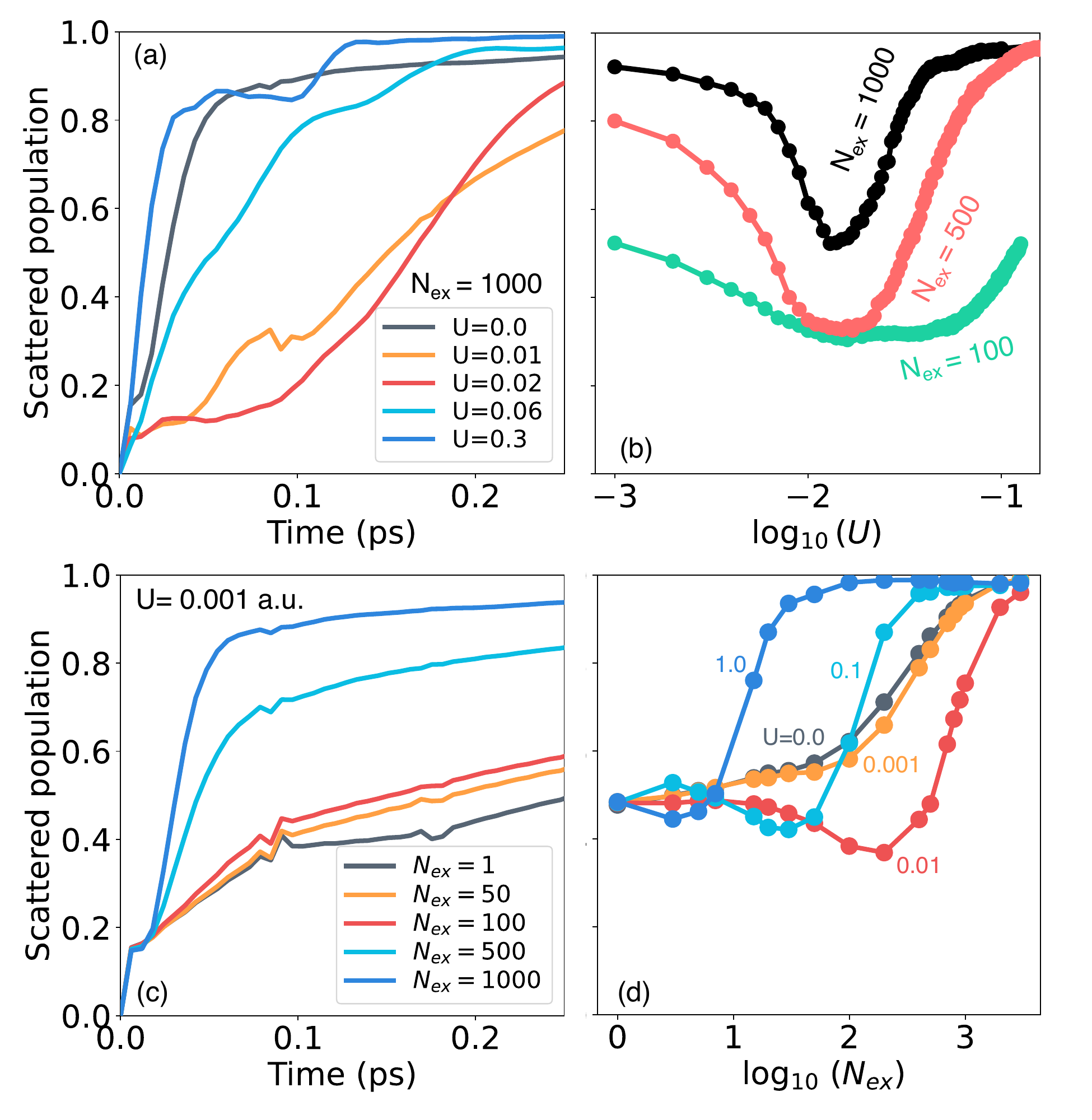}
    \caption{ Scattered population of exciton-polaritons. (a) Time-dependent scattered population at a fixed excitation number ($N_\mathrm{ex} = 1000$) and different on-site interactions. (b)  Scattered population at  $t=181$ fs at various $U$, for $N_\mathrm{ex} = 100$ (green solid line), $N_\mathrm{ex} = 500$ (red solid line) and $N_\mathrm{ex} = 1000$ (black solid line). (c) Time-dependent scattered population  at a constant   $U=0.001 $ a.u. with varying number of excitations $N_\mathrm{ex}$. (d) Scattered population over various numbers of excitations  at  $t=242$ fs, and different on-site interactions. } 
    \label{Fig4}
\end{figure}

In Fig.~\ref{Fig3}, we investigate the exciton-polariton dynamics in the presence of on-site interactions. As observed in Fig.~\ref{Fig1}, the introduction of small on-site repulsive interaction enhances transport. This is illustrated in  Fig.~\ref{Fig3}a, which presents MSD for different on-site interactions and number of excitations. At high $N_\mathrm{ex} = 1000$, the MSD is enhanced at $U = 0.06$ (green solid line) when comparing to the single excited subspace dynamics (black dashed line). However, further increase in $U$, leads to sub-diffusive transport, as illustrated in Fig.~\ref{Fig3}b. Our numerical results reveal that short-time ($t < 50$ fs) ballistic-like dynamics of exciton-polaritons are overall enhanced when the on-site interaction increases. However, at longer times, the MSD and the diffusion constant nonlinearly depend on the on-site interaction strength. 

Fig.~\ref{Fig3}c presents a snapshot of the spatially resolved polariton density at \(t \approx 0.25\) ps at \(U=0\) (blue solid line) and $U=0.02$ (red solid line). Clearly, the polaritonic dynamics at $U=0.02$ is more {\it wavelike} (coherent) than in the absence of on-site interactions. This illustrates that on-site repulsive interactions can counteract phonon-induced dynamic disorder and decoherence and lead to coherent dynamics at high excitations. 

The nonlinear impact of on-site interactions on exciton-polariton dynamics is further illustrated in Fig.~\ref{Fig3}d-f, which presents the time-dependent polariton density at various on-site interaction strengths and at a fixed number of excitations $N_\mathrm{ex} = 1000$. In the absence of on-site interactions, the polaritonic density propagates diffusively. Introduction of a relatively small on-site interaction ($U =$ 0.02 a.u.) leads to the  exciton-polariton density exhibiting ballistic motion with two wavefronts moving in opposite directions (see Fig.~\ref{Fig3}e). In contrast, at much higher on-site interactions, the dynamics of the exciton-polariton density become more complicated, with densities reappearing at the center while the wavefronts expand nonlinearly.

\begin{figure}[t]
    \centering
    \includegraphics[width=1.0\linewidth]{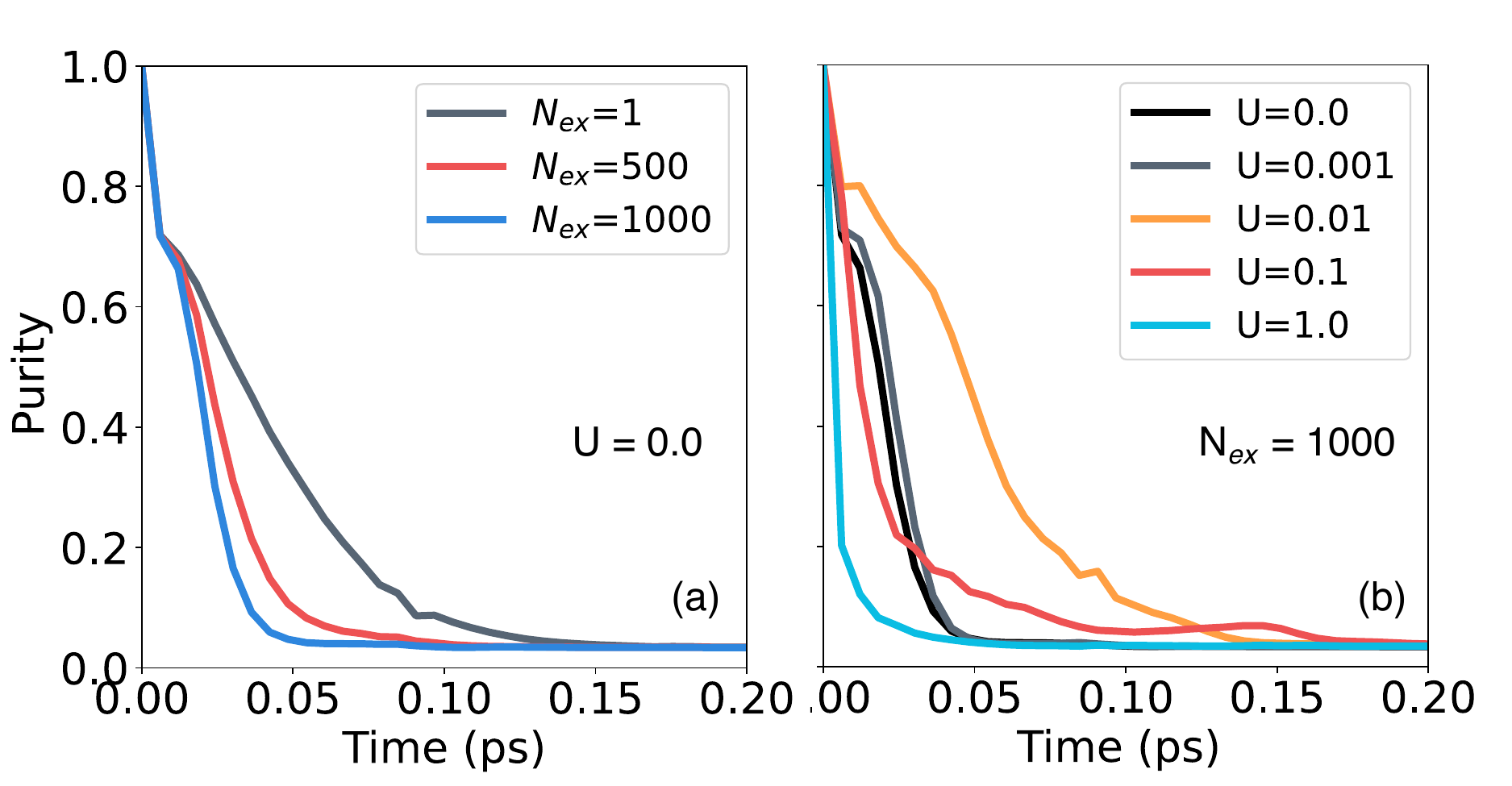}
    \caption{Time-dependent purity of exciton-polaritons. (a) Purity versus time for a single excitation (grey solid line), at $N_\mathrm{ex} = 500$ (red solid line) and at $N_\mathrm{ex} = 1000$ (blue solid line), $U$ is set to 0. (b) Purity versus time at $N_\mathrm{ex} = 1000$  and different on-site interactions.}
    \label{Fig5}
\end{figure}

In Fig.~\ref{Fig4} we analyze how polariton scattering is modified at high excitation subspace. Fig.~\ref{Fig4}a presents the time-dependent scattered population at various on-site interaction strengths. We find that for the excitation chosen here ($N_\mathrm{ex} = 1000$) interactions around $U = 0.02$ minimize polaritonic Fr\"{o}hlich scattering. 

This can be seen more clearly in Fig.~\ref{Fig4}b, which presents the scattered population as a function of interaction strength $U$ at $t = 181$ fs and at various excitation number $N_\mathrm{ex}$. Overall, the scattered populations pass through a minimum (near $U \approx 10^{-2}$ a.u.) where exciton-polariton dynamics also become most coherent.  
We find similar dips in the scattered population (at $t = 181$ fs), for lower $N_\mathrm{ex}$ values (500 and 100). We remind the reader that it is expected (as shown in Fig.~\ref{Fig2}c) that the scattered population rises with increasing $N_\mathrm{ex}$.  Consistent with this expectation, we observe the same upward trend in the scattered population of interacting exciton‑polaritons, as $N_\mathrm{ex}$ increases. This is because higher $N_\mathrm{ex}$ in the system effectively enhances nuclear backreaction, leading to higher phonon-induced dynamical disorder. The interaction strength-dependent scattered population in Fig.~\ref{Fig4}b is relatively flatter at lower excitations (e.g. $N_\mathrm{ex} = 100$).   The reason for this flattening is twofold. One, at low $N_\mathrm{ex}$, the overall scattering induced by the phonon is significantly lower, and two, the effective non-linear interaction is also much less (scaled by $N_\mathrm{ex}$). 
Conversely, at higher excitations ($N_\mathrm{ex} \gg 100$) the impact of $U$ is more pronounced, leading to a clear dip in the interaction strength-dependent scattered population. 

In Fig.~\ref{Fig4}c–d, we examine the dependence of polariton scattering on \(N_{\rm ex}\). In the weak on-site interaction limit (\(U = 0.001\) a.u.), the scattered population grows monotonically, consistent with the behavior shown in Fig.~\ref{Fig2}. Interestingly, under stronger on-site interactions, the scattered population develops a clear minimum as a function of \(N_{\rm ex}\) as shown in Fig.~\ref{Fig4}d. This non-monotonic trend reveals that there is an optimal excitation number at which Fr\"{o}hlich scattering is most effectively suppressed, leading to enhanced polariton coherence. Further, we also find that with increasing on-site interaction, the optimal excitation number gradually shifts to the left, while the curve around the minima also becomes shallower (while vanishing for $U \ll$ 0.01). 

In Fig.~\ref{Fig5} we present the purity of the exciton-polariton density, which quantifies the amount of coherence in the exciton-photon subsystem. A purity of unity corresponds to a pure state, and its decay corresponds to decoherence. Fig.~\ref{Fig5}a compares the time-dependent purity at three different excitations, $N_\mathrm{ex} = 1$, 500 and 1000 in the absence of on-site excitonic interaction. We find that the decay in purity monotonically increases with the increase in $N_\mathrm{ex}$, which corroborates our observation in Fig.~\ref{Fig2}. This enhancement in the decay of purity is due to the enhancement of phonon-induced dynamical disorder and Fr\"{o}hlich scattering of polaritons. Meanwhile, Fig.~\ref{Fig5}b presents the time-dependent purity at various on-site interaction strengths. We find that the purity decays the slowest when choosing $U = 0.01$ which coincides with the dip in the scattered population in Fig.~\ref{Fig4}b. Therefore, as mentioned above,  a particular range of optimal on-site  interaction strength (which, in this case, is around $U \sim 10^{-2}$) the exciton-polariton remains most coherent. Overall, the time-dependent purity supports our understanding of how on-site repulsive interactions and excitation-induced dynamical disorder affect the exciton-polariton dynamics.

\section{Conclusions} 
In this work, we introduce a mean-field many-body Ehrenfest approach that seamlessly combines the multitrajectory Ehrenfest method, which propagates phonons quasi-classically, with a mean-field propagation of the excitonic–photonic many-body wavefunction beyond the single excited subspace. Our simulations beyond the single-excitation subspace illustrate the interplay between phonon-induced dynamic disorder and on-site repulsive exciton-exciton interactions in both bare excitonic materials and strongly coupled exciton-polariton systems. Importantly, we find a non-monotonic dependence of polariton scattering and decoherence on both on-site interaction strength and excitation number. We show that, intermediate on-site interactions can suppress phonon-induced decoherence, leading to optimal coherence at a specific combination of on-site interaction strength and excitation number. Our purity analysis further corroborates these findings, revealing maximal coherence lifetimes under these optimal conditions.

In addition to providing the microscopic insights into the exciton-polariton transport beyond the single excited subspace, our work lays a crucial foundation for the future development of more accurate yet computationally tractable quantum-dynamical methods. Our future plans include extending the mean-field Gross-Pitaevskii description to systematically include correlations~\cite{AlonPRA2008} as well as using more accurate mixed quantum-classical approaches~\cite{HuoJCP2011, Shakib2017JPCL, Cotton2019JCP,Cotton2019JCP2, MandalJCTC2018, Cotton2019JCP, Mannouch2023JCP, Mannouch2020JCP} to propagate nuclear dynamics. 
\\

\section{Acknowledgments}
This work was supported by the Texas A\&M startup funds. This work used TAMU FASTER at the Texas A\&M University through allocation  PHY230021 from the Advanced Cyberinfrastructure Coordination Ecosystem: Services \& Support (ACCESS) program, which is supported by National Science Foundation grants \#2138259, \#2138286, \#2138307, \#2137603, and \#2138296.  The authors appreciate discussions with Braden Weight, Amir Amini, and Logan Blackham.  A.M. is deeply thankful for the inspiring discussions with the late Hendrik Monkhorst.

\bibliography{bib.bib}
\bibliographystyle{naturemag}

\end{document}